\documentclass[aps,prd,preprintnumbers,nofootinbib,floatfix,longbibliography,notitlepage,11pt]{revtex4-1}

\usepackage{amsfonts,amsmath,amssymb,yfonts,graphicx,epsfig,xcolor,yhmath} 
\usepackage{hyperref}
\usepackage{url}

\newcommand{\be}{\begin{equation}}
\newcommand{\ee}{\end{equation}}
\newcommand{\ba}{\begin{eqnarray}}
\newcommand{\ea}{\end{eqnarray}}
\newcommand{\nn}{\nonumber}

\newcommand{\qk}{q_{\vec{k},s}}

\makeatletter
\long\def\dddddot#1{%
  {\mathop {#1}\limits ^{\vbox to-1.4\ex@ {\kern -\tw@ \ex@ \hbox {\normalfont .....}\vss }}}%
}
\long\def\multidots#1#2{%
  \count@=0
  {{\mathop {#2}\limits ^{\vbox to-1.4\ex@ {\kern -\tw@ \ex@ \hbox {\normalfont %
  \loop%
  \ifnum#1>\count@%
  .%
  \advance\count@ by1%
  \repeat%
  }\vss }}}}%
}
\makeatother

\begin{document}

\title{Quantum gravity corrections to the fall of the apple}
\author{Samarth Chawla$^{1}$ and Maulik Parikh$^{1,2}$}
\affiliation{$^{1}$Department of Physics, Arizona State University, Tempe, Arizona 85287, USA}
\affiliation{$^{2}$Beyond: Center for Fundamental Concepts in Science, Arizona State University, Tempe, Arizona 85287, USA} 
\begin{abstract}
\begin{center}
{\bf Abstract}
\end{center}
\noindent
We consider the motion of a massive particle in a static, weakly-curved spacetime where the gravitational field is taken to be quantized. We find that Newton's law of free-fall is modified by quantum-gravitational corrections, in addition to the known special-relativistic and post-Newtonian modifications. The quantum-gravitational corrections take the form of stochastic noise in the particle trajectory, where the statistical properties of the noise depend on the quantum state of the gravitational field.
\end{abstract}

\maketitle

\section{Introduction}

\noindent
In 1666, while England was locked down during an epidemic of the plague, a young Isaac Newton spent his quarantine rather fruitfully, wandering about apple orchards. There, legend has it that, occasioned by the fall of an apple, he discovered his universal law of gravitation. Of course, for all its successes, Newton's law is incomplete. We know from special relativity that Newtonian free-fall admits corrections in powers of $v^2/c^2$, and we know from general relativity that there are post-Newtonian corrections which go roughly as powers of $GM/r$. We can also anticipate quantum corrections, once gravity is treated as an effective field theory. Such corrections would be expected on dimensional grounds to go as powers of $\frac{\hbar G}{c^3 r^2}$. Indeed, a famous calculation ~\cite{Duff:1974ud,Donoghue:1993eb,Donoghue:1994dn} of the one-loop quantum-gravitational modification to the Newtonian potential, analogous to radiative corrections to Coulomb's law in quantum electrodynamics, predicts a correction of exactly this form; several other authors have performed similar calculations, though there are some discrepancies in the details ~\cite{Muzinich:1995uj,Hamber:1995cq,Akhundov:1996jd,Khriplovich:2002bt,Bjerrum-Bohr:2002gqz}.

The one-loop correction to the Newtonian potential is exceedingly small $\sim \left ( \frac{l_P}{r} \right )^{\! 2}$ and is surely unmeasurable. However, loops are not the only way that the quantization of gravity can be manifested: there can also be tree-level effects arising from the existence of very unclassical quantum states. That is, if the quantum state of the gravitational field is (loosely speaking) very far from being a coherent state, there could be tree-level processes that cannot be understood within the context of classical gravity. Consider by analogy the experimental evidence for the quantization of light. This comes not just from loop effects as in the Lamb shift, but also from tree-level phenomena such as the Compton effect, photon anti-correlation, and sub-Poissonian photon statistics ~\cite{Clauser:1974gd,PhysRevLett.39.691}. In these instances the quantum state of the radiation field is often an eigenstate of the number operator or perhaps a squeezed state i.e. some state that does not correspond to a field configuration of classical electrodynamics. 

Similarly, we can look for signatures of quantum gravity that are pronounced in special states of the gravitational field. It is important to emphasize that just because we observe a world in which gravity appears to behave classically does not mean that the gravitational field is in a coherent state. While it is true that the expectation value of the field operator in a coherent state is the classical field configuration, there are infinitely many other states that are not coherent states for which that is also true. To see this, suppose we are given a unitary displacement operator $D_{\alpha} = \exp(\alpha a^\dagger - \alpha^* a)$ ~\cite{Glauber:1963tx}. Then, because $[a,D_{\alpha}] = \alpha D_{\alpha}$, the expectation value of the field mode operator $\phi \sim a + a^\dagger$ is the same in the state $D_{\alpha} | \Psi \rangle$ as in the coherent state $D_\alpha | 0 \rangle$, for any arbitrary state $|\Psi \rangle$ that satisfies $\langle \Psi | \phi | \Psi \rangle = 0$. Naturally arising non-coherent states of the gravitational field include thermal states and squeezed states. Thermal states of the gravitational field could originate from evaporating black holes or from the cosmic gravitational wave background. Squeezed gravitational states could be produced by inflation ~\cite{Grishchuk:1990bj,Albrecht:1992kf,Koks:1996ga} and even more simply by ordinary classical sources nonlinearly coupled to gravity.
 
Recently the implications of the quantization of gravity on the geodesic deviation between two free-falling masses were analyzed ~\cite{Parikh:2020nrd,Parikh:2020kfh,Parikh:2020fhy}. It was found that quantization of the spacetime metric induces noise (random fluctuations) in the geodesic separation of pairs of falling particles. The statistical properties of the noise -- its standard deviation, autocorrelation function, etc. -- depend on the quantum state of the gravitational field; for certain states, notably squeezed states, the noise amplitude is greatly enhanced so that it might even be possible for this fundamental noise to be experimentally observable in the fluctuations of the mirror separation of gravitational wave detectors. Mathematically, in place of the deterministic geodesic deviation equation we now have a stochastic Langevin-like equation. This is indeed a general phenomenon: as Feynman and Vernon showed ~\cite{Feynman:1963fq}, when two quantum systems are coupled, integrating out one system causes the second system's dynamics to become stochastic even in its classical limit. Intuitively, that is because the final state of the system that was integrated out must be summed over.

Here we will apply the formalism developed in ~\cite{Parikh:2020nrd,Parikh:2020kfh,Parikh:2020fhy} to the case of Newtonian free-fall. Instead of considering the geodesic deviation of two particles in flat space, we will consider a single particle falling in the background of a weak, static gravitational field, as seen by an observer at rest with respect to the background. After taking some care with fixing the gauge, the perturbations of the gravitational field will be treated quantum-mechanically. As we will see, the quantum aspects of the computation map directly to the previous case. Integrating out the quantized gravitational field and taking the classical limit for the particle yields its equation of motion. We find (\ref{eom}) that Newton's law is augmented by special-relativistic, post-Newtonian, and quantum-gravitational corrections. In particular, there are random fluctuations in the force on the particle, the statistical properties of which depend on the quantum state of the gravitational field. Thus Newton's apple would not fall straight down but would be subject to minute quantum jitters, which can be thought of heuristically as arising from the random bombardment of the apple by gravitons. The scale of the fluctuations depends on the quantum state but is likely to be too small to measure.

\section{Setup}

We are interested in quantum-gravitational corrections to free-fall; we will have in mind Newton's apple falling in the earth's gravitational field as seen by a terrestrial observer. (We will neglect not only air resistance but also the rotation of the earth.) Before taking into account quantum effects, let us consider the dynamics of a massive, nonrelativistic particle in a vacuum region of a weakly-curved static spacetime. Since the spacetime is static, we can write the line element as
\be 
\label{background metric}
ds^{2} = -(1 + 2 \phi + 2 \phi^2 + 2 \psi+...) dt^{2} + ((1 - 2\phi) \delta_{ij} +
g^{(4)}_{ij}+ ...)dx^i dx^j,
\ee
where $\partial_t$ is the timelike Killing vector, none of the metric components depends
on $t$, and we have $g_{0i} = 0$. Here $\phi^2$, $\psi$, and $g^{(4)}_{ij}$ are the leading post-Newtonian corrections to the metric. 

The linearized Einstein equations consist, to leading order, of the single equation
\begin{equation}
-\nabla^{2} \phi = 4 \pi G T_{00},
\end{equation}
which is the Poisson equation for a Newtonian potential $ \phi$. The geodesic equation in the nonrelativistic limit $ \dot t \sim 1$, $\dot x^{i} \ll 1$ gives Newton's universal law of gravitation:
\be   
\vec{F} = m \ddot{\vec{x}} = - m \vec{\nabla} \phi.
\label{Newton}
\ee
Our goal is to find the modification to this law, when the spacetime metric is quantized.

The coordinates chosen to write the line element are a convenient choice to exhibit the
static nature of the background. They do not, however, form the proper reference frame of an observer at rest on the ground. The proper reference frame of such an observer is an orthonormal tetrad ${e^{\mu}_{0}, e^{\mu}_{j}}$ at the observer's worldline, with the timelike basis vector always equaling the observer's four-velocity: $e^{\mu}_{0} = u^\mu$. The extension of such a tetrad to a local coordinate system is naturally generated by the
exponential map on the spacelike members of the observer's tetrad. Each such geodesic can be labeled by the proper time of the observer at its initial point. This naturally becomes the time coordinate in these coordinates.

For a freely falling observer this construction results in Fermi-normal coordinates. A
static observer (that is, one following an integral curve of the timelike Killing vector
field) is not on a geodesic however, and his or her nonzero acceleration results in a slight
modification to the evolution of the tetrad along the observer's worldline. Instead of the
tetrad being parallel transported, it undergoes Fermi-Walker transport:
\begin{equation} \label{} 
    u^{\mu} \nabla_{\mu} e^{\nu}_{b} = (u^{\nu}a_{\sigma} - a^{\nu}u_{\sigma})
    e^{\sigma}_{b},
\end{equation}
where Latin letters represent tetrad indices and $ a^{\mu}$ is the rate of change of $
u^{\mu}$ relative to parallel transport. The local coordinate system thus constructed $ (\tau, \xi^{i})$ has a few notable features:
\begin{itemize}
    \item $ \sqrt{\xi_{i}\xi^{i}} $ is the proper distance of the point $ (\tau, \xi^{i})$
        from the observer $ (\tau, 0)$.
    \item The only nonzero first derivative of the metric at $ \xi^{i} = 0$ is $
        \partial_{\mu}g_{\tau \tau} = -2 a_{\mu}$.
    \item The second derivatives of the metric at $ \xi^{i}=0$ can be written in terms of
        the acceleration and components of the Riemann tensor.
\end{itemize}
The line element in these coordinates reads
\begin{multline} \label{} 
    ds^{2} = -(1 + 2 a_{j}\xi^{j} + 2 a_{j}a_{k} \xi^{j}\xi^{k} + R_{\tau j \tau k}
    \xi^{j}\xi^{k})d\tau^{2} + \frac{4}{3} R_{\tau jkl}\xi^{j}\xi^{k} dx^{l}d\tau
\\
+ \left(\delta_{jk} + \frac{1}{3} R_{mjkn} \xi^{j}
\xi^{k}\right) dx^{m}dx^{n} + O(\xi^{3}),
\end{multline}
where $ a^{\mu}$ and $ R_{\mu \nu \rho \lambda}$ are evaluated at the origin. When $a^\mu = 0$, one can confirm that this reduces to the metric in Fermi normal coordinates.

\subsection{Gauge-fixing the metric perturbation}
Let us consider perturbations about a given background
    \be
    g_{\mu \nu} = g^{(0)}_{\mu \nu} + h_{\mu \nu},
    \ee
where $g^{(0)}$ solves Einstein's equations sourced by a static, fixed $T_{\mu \nu}$ (i.e.,  contains post-Newtonian corrections to all orders) and $h$ solves linearized vacuum Einstein's equations on this curved background. The gauge conditions $\nabla^{\mu} h_{\mu \nu} = 0 = h$ are a valid choice for vacuum perturbations in regions where the Ricci tensor of the background vanishes, which happens everywhere outside the source. In this gauge, the linearized Einstein equations are
    \be \label{curved background linearized Einstein}
        \frac{1}{2} \nabla^{\alpha}\nabla_{\alpha} h_{\mu \nu} + R_{\rho \mu \sigma
        \nu}h^{\rho \sigma} = 0,
    \ee
and there is residual gauge freedom parametrized by vector fields $ \xi^{\mu}$ that satisfy $ \nabla_{\alpha}\nabla^{\alpha} \xi^{\mu} = 0 = \nabla_{\alpha}\xi^{\alpha}$. For a flat background, one fixes this residual gauge freedom by imposing $ h_{j0}=0$, which (along with the equations of motion) implies $ h_{00}=0$.

Such a gauge choice is not always possible on a curved background. The equation of motion for $h_{j 0}$ is $\nabla_{\alpha} (\nabla^{\alpha} h_{\mu \nu}) e^\mu_j e^\nu_0 + 2 R_{\rho j \sigma 0} h^{\rho \sigma} = 0$. This in general includes terms involving $h_{i j}$ and $\partial_{\nu}h_{ij}$. If the condition $h_{j0} = 0$ were to hold, the equation of motion for $h_{j 0}$ would impose a nontrivial constraint on $h_{ij}$, but there are not enough degrees of freedom left in $\xi^\mu$ to both satisfy the constraint and make $h_{j 0}$ vanish. Said another way, the obstruction to choosing a gauge where $ h_{j0}=0$ is the presence of inhomogeneous terms in the equation of motion for $ h_{j0}$. However, for a static background the inhomogeneous terms are at most ${\cal O}(\partial_{j}\phi \partial_{k} \phi)$. Thus, there is a consistent gauge choice where $ h_{j0}$ is also at most ${\cal O}(\partial_{j}\phi \partial_{k}\phi)$, and is thus subleading when compared to $h_{jk}$.

In flat space, it is argued that since the vanishing of $ \partial^{\mu} h_{\mu \nu}$ renders $h_{00}$ non-dynamical, and the vanishing of $ h_{j0}$ makes its equation of motion a Laplace equation, $ h_{00}$ vanishes for suitable boundary conditions. From the preceding discussion it is no surprise then, that a similar argument goes through at leading order, and subleading corrections to $ h_{00}$ appear only at ${\cal O}(\partial_{j}\phi\partial_{k}\phi)$.  Thus, to leading order, we will still be able to work in the transverse-traceless (TT) gauge:
    \begin{align} \label{} 
        \nabla^{\mu}h_{\mu \nu} &= 0, \\
        g^{\mu \nu} h_{\mu \nu} &= 0,\\
        h_{0 \mu} &= 0 + O((\partial_{k}\phi)^{2}).
    \end{align}
  
\subsection{Action}
The action for a freely falling point particle of mass $m_0$ is 
\begin{equation} \label{pp action} 
    \begin{split}
        S_{pp} &= -m_{0}\int \sqrt{- ds^{2}} \\
               &= -m_{0}\int d\tau \Bigg[(1 + 2 a_{j}\xi^{j} + 2 a_{j}a_{k} \xi^{j}\xi^{k} +
               R_{0 j 0 k} \xi^{j}\xi^{k}) + \frac{4}{3} R_{0 jkl}\xi^{j}\xi^{k}
           \dot{\xi}^{l} \\
               &\ \ + \left(\delta_{jk} + \frac{1}{3} R_{mjkn} \xi^{j} \xi^{k}\right)
       \dot{\xi}^{m} \dot{\xi}^{n} + O(\xi^{3}) \Bigg]^{ \frac{1}{2}}. 
    \end{split}
\end{equation}
where we have inserted the metric as written in the observer's reference
frame. Since we have in mind non-relativistic, Newtonian free fall, we will be expanding the square root order by order in the three parameters $ \dot{\xi}, \phi, h$. All three are
independent of each other, and we will keep the first order term in each beyond the
standard Newtonian non-relativistic action. Since $ a_{j}$ and the Riemann components above are to be evaluated at the origin, we can
use the coordinates introduced in \eqref{background metric} and make the appropriate basis
transformation to the observer's tetrad. In fact, since the observer's four-velocity is
precisely $ \partial_{t}$ and we have the freedom to choose a spacelike tetrad at the
origin, we can align the observer's tetrad with the $ (t, x^{i})$ coordinate basis.

Keeping the leading order in each correction,
\begin{equation} \label{} 
    S_{pp} =  m_{0} \int dt \left ( \frac{1}{2}\dot{\xi}^{2} + \frac{1}{8} \dot{\xi}^{4} - \phi 
    -\frac{1}{2} \phi^{2} - \frac{3}{2} \phi \dot{\xi}^{2} - \psi + \frac{1}{4}
\ddot{h}_{jk} \xi^{j} \xi^{k} \right ).
\end{equation}
This point particle action contains the leading post-Newtonian term at ${\cal O}(\phi^{2}, \psi)$, the leading special-relativistic correction at ${\cal O}(v^{4}, \phi v^{2})$ and the leading contribution from the perturbation at ${\cal O}(h)$.

We also need the action for linearized gravity.
In TT gauge the Einstein-Hilbert action about an arbitrary vacuum background reduces to
\begin{equation} \label{Lorenz gauge Einstein-Hilbert action} 
    S_{EH} = \frac{1}{64 \pi G} \int d^{4}x \left(h_{jk} \nabla_{\alpha}\nabla^{\alpha}
    h^{jk} - 2 h_{jk} R^{jmkn} h_{mn} \right).
\end{equation}
For a weakly curved background, the leading action in the chosen gauge is of course identical to the flat space action, with corrections at ${\cal O}(h^{2} \partial_{j} \phi )$:
\begin{equation} \label{leading linearized Einstein-Hilbert action} 
    S_{EH} = \frac{1}{64 \pi G} \int d^{4}x \left(h_{jk} \square h^{jk} \right).
\end{equation}
Adding together the linearized Einstein-Hilbert action and the point particle action, and
integrating by parts to remove second derivatives in time,
\begin{multline} \label{} 
    S = -\frac{1}{64 \pi G} \int d^{4}x \left(\partial_{\alpha}h_{jk} \partial^{\alpha}
    h^{jk} \right) \\
    + m_{0} \int dt \left ( \frac{1}{2}\dot{\xi}^{2} + \frac{1}{8} \dot{\xi}^{4} - \phi 
    -\frac{1}{2} \phi^{2} - \frac{3}{2} \phi \dot{\xi}^{2} - \psi - \frac{1}{2}
    \dot{h}_{jk} \dot{\xi}^{j} \xi^{k} \right ).
\end{multline}
We can break $ h_{ij}$ into a sum over discrete modes within a box of volume $V$
\be
{h}_{ij}(t,\vec{x}) = \frac{1}{\sqrt{\hbar G}} \sum_{\vec{k},s} \qk(t) e^{i \vec{k} \cdot \vec{x}} \epsilon^s_{ij} (\vec{k})\ . 
\label{hdecomp}
\ee
where $q$ is the mode amplitude, and $\epsilon$ is the polarization tensor.
Focusing on a single mode of frequency $\omega$ and dropping a polarization \cite{Parikh:2020fhy}, we find
\be
L = m_0 \left (\frac{1}{2} \dot{z}^2 + \frac{1}{8} \dot{z}^4 - \frac{3}{2} \phi \dot{z}^2
- \phi - \frac{1}{2} \phi^2 - \psi \right ) + \frac{1}{2} m (\dot{q}^2 - \omega^2 q^2) - g
\dot{q} \dot{z} z,
\label{Lagrangian}
\ee
where we have relabeled the particle's position as $z(t)$ and defined $m \equiv \frac{V}{16 \pi \hbar G^2}$ and $g \equiv \frac{m_0}{2 \sqrt{\hbar G}}$. This Lagrangian describes a particle derivatively coupled to a harmonic oscillator, $q(t)$. The terms we have omitted for simplicity, including the other polarization, can also be calculated \cite{Kanno:2020usf,Cho:2021gvg}.

\section{Quantization}

Having obtained the action, we are ready to quantize the theory. We will at first treat both the particle and the gravitational field quantum-mechanically; after integrating out gravity, we will take the classical limit for the particle. Our approach here is essentially identical to the one developed in ~\cite{Parikh:2020kfh,Parikh:2020fhy}. That approach differs from standard calculations in effective field theory in two significant ways. First, the initial state of the gravitational field will be arbitrary, rather than the vacuum state typically (though not always) chosen in effective field theory; we imagine that there is some initial quantum gravitational state that is given to us by astrophysical sources. Second, as a result of the interaction with the particle, the final state of the particle-field system is typically an entangled state; intuitively, the falling particle can absorb gravitons as well as emit them through spontaneous and stimulated emission. This means in particular that there is no definite final state for the gravitational field, and we will have to sum over final states of the field . Consequently we cannot integrate out gravity by, as it were, doing a path integral over gravitational field configurations to obtain an effective theory for the particle. Instead, the best we can do is to determine the probabilities (not amplitudes) for the particle to be in various final states. The problem of integrating out a system for which the final state is summed over was treated in detail by Feynman and Vernon ~\cite{Feynman:1963fq}, and its application to stochastic gravity was studied by Hu and collaborators ~\cite{Calzetta:1993qe,Hu:1999mm,Johnson:2000if}.

Suppose then that the gravitational field is initially in a state $|\Psi \rangle$. The object of interest is the transition probability for the particle to go from state $A$ to state $B$ in some time $T$. Here $A$ and $B$ could correspond to wavepackets localized in position, but as we will see the dynamics will not depend on what $A$ and $B$ actually are. We have
\be
P_\Psi (A \to B) = \sum_{|f \rangle} | \langle f, B | \hat{U}(T) | \Psi, A \rangle|^2,
\ee
where $U$ is the unitary time-evolution operator that can be obtained from the Hamiltonian. Inserting a complete set of position eigenstates, we have
\begin{multline}
P_\Psi (A \to B) = \int dz_i d z'_i dz_f dz'_f \phi^*_A(z'_i) \phi_B (z'_f) \phi_B^* (z_f)
\phi_A (z_i)  \\
\times \sum_{|f \rangle} \langle \Psi, z'_i | \hat{U}^\dagger(T) | f, z'_f \rangle \langle
f, z_f | \hat{U}(T) | \Psi , z_i \rangle,
\label{prob}
\end{multline}
where $\phi_A (z)$ is the position-space wavefunction in the state $A$, etc. Taking the Legendre transform of (\ref{Lagrangian}), we find that the Hamiltonian is
\begin{multline}
H(q,p,z,\pi)=\left(\frac{p^2}{2m}+\frac{\pi^2/2 + gp\pi z/m}{m_0 (1 - 3
\phi(z))}\right)\left(1-\frac{g^2 z^2}{mm_0 (1-3 \phi(z))}\right)^{\!
-1}+\frac{1}{2}m\omega^2q^2 \\
+ m_0 \left (\phi +  \frac{1}{2} \phi^2 + \psi \right ) .\label{fullHam}
 \end{multline}
where $\pi$ and $p$ are the canonical momentum conjugates of $z$ and $q$ respectively. Note that in writing (\ref{fullHam}) we have discarded the relativistic $\dot{z}^4$ term in (\ref{Lagrangian}), for simplicity. Using the Hamiltonian we can express each of the amplitudes in (\ref{prob}) in canonical path-integral form:
\be
\langle q_f , z_f | \hat{U}(T) | q_i , z_i \rangle = \int {\cal D} \pi {\cal D} z {\cal D} p {\cal D} q \exp \left (\frac{i}{\hbar} \int_{0}^{T} dt \left ( \pi \dot{z} + p \dot{q} - H(q,p,z,\pi) \right ) \right )\,.
\ee
Performing the Gaussian path integral over $\pi$ yields
\begin{multline}
\langle q_f , z_f | \hat{U}(T) | q_i , z_i \rangle = \int \tilde{\cal D} z
e^{\frac{i}{\hbar} \int dt m_0 \left (\frac{1}{2} \dot{z}^2 - \frac{3}{2} \phi \dot{z}^2 -
\phi - \frac{1}{2} \phi^2 - \psi \right )} \\
\times \int {\cal D} p {\cal D} q \exp \left (\frac{i}{\hbar} \int_{0}^{T} dt \left (p \dot{q} - H_z(q,p) \right ) \right )\ ,
\label{amppi}
\end{multline}
where $\tilde{\cal D} z$ is a measure in which some $z$-dependent pieces have been absorbed. Here we have also defined
\be
H_z(q,p) \equiv \frac{(p + g z \dot{z})^2}{2m} + \frac{1}{2} m \omega^2 q^2 \ , \label{Hz}
\ee
Despite the slightly different setup, this expression is precisely the same as equation (35) in ~\cite{Parikh:2020fhy}, and we can simply follow that derivation to its end. In short, we  compute the path integrals over $p$ and $q$ in (\ref{amppi}) by expressing them as a probability amplitude that is easiest to evaluate canonically, then sum over the final states in (\ref{prob}), and finally integrate over all modes. Having done so, we will have completely integrated out the gravitational degrees of freedom, leaving a probability expressed entirely in terms of the particle degree of freedom, $z$:
\be
P \sim \! \! \int \tilde{\cal D} z \tilde{\cal D} z' e^{\frac{i}{\hbar} \int dt m_0 \left
(\frac{1}{2} \dot{z}^2  - \frac{3}{2} \phi \dot{z}^2 - \phi - \frac{1}{2} \phi^2 - \psi
\right )} e^{-\frac{i}{\hbar} \int dt m_0 \left (\frac{1}{2} \dot{z'}^2 - \frac{3}{2}
\phi' \dot{z'}^2 - \phi' - \frac{1}{2} \phi'^2 - \psi' \right )}F_{\Psi} [z,z'],
\label{zprob}
\ee
where primed functions have $z'$ as their argument: $\phi' = \phi(z')$ etc. The above expression is a double path integral because it corresponds to a probability rather than to an amplitude. Note that exponents are complex conjugates of each other. But the probability does not factorize into an amplitude times its conjugate because of the presence of $F_{\Psi}$, the Feynman-Vernon influence functional. This encodes the entirety of the effect of the quantum gravitational field, which recall was initially in the state $|\Psi \rangle$, on the particle degree of freedom.

For many interesting classes of states (the vacuum, coherent states, thermal states, squeezed states), the influence functional can be computed exactly ~\cite{Parikh:2020fhy}. Now, as shown by Feynman and Vernon~\cite{Feynman:1963fq}, a generic feature is that the absolute value of the influence functional can be written in a very suggestive form. Using an identity that is essentially the infinite-dimensional generalization of $e^{\frac{b^2}{4a}} = \int dy e^{-a y^2 + by}$, we can write the absolute value of $F$ as a statistical average:
\ba
|F_{\Psi}| = &&\exp\left[-\frac{m^2_0}{32\hbar^2}\int_0^T\int_0^T dt\,dt'\,A_{\Psi}(t,t')\left(X(t)-X'(t)\right)\left(X(t')-X'(t')\right)\right] \nn\\
&& 
\hspace{-7mm}
 = 
\int {\cal
D}N\exp\left[-\frac{1}{2}\int_0^T\int_0^Tdt\,dt'\,A_{\Psi}^{-1}(t,t')N(t)N(t')+\frac{i}{\hbar}\int_0^T
dt\frac{m_0}{4}N(t)\left(X(t)-X'(t)\right)\right]\hspace{-1mm},
\label{funcgauss}
\ea
where $X = \frac{d^2}{dt^2} (z^2) $ and $X' = \frac{d^2}{dt^2} ({z'}^2)$.
The right-hand side can be interpreted as a statistical average over a random function $N(t)$, with a zero-mean Gaussian probability distribution. Thus $N(t)$ is a noise function whose statistical properties are encoded in the auto-correlation function $A_\Psi(t,t') =  \langle \langle N_{\Psi} (t)N_{\Psi} (t') \rangle \rangle$. This can be explicitly evaluated for many interesting classes of states. For the vacuum, we find
\be
A_0(t,t') = \frac{4\hbar G}{\pi}\int_0^\infty d\omega\,\omega\cos(\omega(t-t')).
\label{A0}
\ee
This is formally divergent but can be regulated by measurement sensitivity cut-offs. The power spectrum of the fluctuations, $S(\omega) = \int dt e^{-i \omega t} A(t,0)$, can also be evaluated. For the vacuum, we find $S_{\rm vac} (\omega) = 4 G \hbar \omega$ while for a uniformly squeezed state with squeezing parameter $r$, we have $S_{\rm squeezed} (\omega) \sim 4 G \hbar \omega (\cosh(2r))$ ~\cite{Parikh:2020fhy}; it is this exponential enhancement in the noise for squeezed states that makes them so phenomenologically interesting ~\cite{Hertzberg:2021rbl}.

Inserting (\ref{funcgauss}) into (\ref{zprob}) (we neglect the phase of $F_{\Psi}$, which is responsible for radiation reaction effects) we find that the transition probability is a triple path integral over exponentials. Two of the path integrals, over $z$ and $z'$, are the usual ones that appear in a quantum-mechanical probability. By taking a saddle point over these (and demanding that the two resulting equations be the same), we can take a classical limit. We find the generalization of Newton's law:
\be
\ddot{z} = - \frac{d \phi }{dz}+ \frac{1}{2} \phi \frac{d \phi}{dz} - \frac{d \psi }{dz} + \ddot{N}_{\Psi} z.
\label{eom}
\ee
But note that this still leaves a third path integral, over $N$. That remaining path integral indicates the presence of a statistical average.

\section{Discussion}

We can express our equation of motion in vector form:
\be
\vec{F} = m \ddot{\vec{x}} = \underbrace{- m \vec{\nabla} \phi}_{\rm Newton} -\underbrace{m \frac{3}{2} \dot{\vec{x}}^2 \ddot{\vec{x}}}_{\rm SR} + \underbrace{\frac{1}{2} m \phi \vec{\nabla} \phi - m \vec{\nabla} \psi}_{\rm GR}  + \underbrace{m \ddot{N}_{\Psi} \vec{x}}_{\rm QG}.
\label{eomvector}
\ee
Here we have reintroduced the relativistic term in the Lagrangian by hand; essentially, we assumed that the relativistic correction does not affect the quantum calculation, to leading  order. Our equation of motion, which generalizes (\ref{Newton}), contains the leading corrections to Newton's law. We have the familiar special-relativistic and post-Newtonian leading corrections. But in addition we also have a noise term, originating in quantum gravity, $\vec{F}_{\rm QG} = m \ddot{N}_{\Psi} \vec{x}$ (more precisely, this force does not have to be aligned with $\vec{x}$ but can involve a noise tensor acting on $\vec{x}$ \cite{Cho:2021gvg}). The noise term, being a random function, has a fundamentally different character from the other terms in the equation of motion: this is now a stochastic differential equation rather than a deterministic one. The statistical properties of $N_\Psi$ depend on $|\Psi \rangle$, the state of the gravitational field; as mentioned, they can be computed exactly for many classes of states.

There is a similar effect in electrodynamics. Consider the classical motion of a point particle interacting with a
quantized electromagnetic field. The Lorentz force law is corrected by radiation reaction classical interactions as
well as a stochastic term coming from the state-dependent quantum fluctuations of the electromagnetic field.
Together, these lead to an Abraham-Lorentz-Langevin equation \cite{Galley:2006gs}. There are, however, some key
differences with the gravitational case. First, the interaction term in the Lagrangian is different in form ($q
\vec{A} \cdot \vec{v}$ versus $m h_{ij} \dot{\xi}^{i} \dot{\xi}^{j}$). Consequently, in the electromagnetic case,
the noise appears in the equation of motion as an inhomogeneous term, akin to a stochastic driving force, whereas
in the gravitational case the noise term appears homogeneously as a stochastic parametric oscillation. Another
difference is that the electromagnetic noise term arises even for an isolated point charge, while in the
gravitational case one has to consider at least a pair of particles even when the gravitational is quantized.

It is interesting that the mass of the particle drops out of the equation of motion (\ref{eom}) even when quantum-gravitational fluctuations are taken into account. Thus the weak equivalence principle continues to hold, at least when the gravitational backreaction of the falling particle is neglected. Now, the principle of equivalence is of course what led Einstein to a geometric view of gravity. That our stochastic free-fall equation does not involve the mass of the falling particle therefore suggests the following interpretation. Each realization of the noise corresponds to the particle falling in a spacetime with a distinct geometry. The probability of each such geometry is given by the corresponding probability for the realization, with the probabilities having a Gaussian distribution. It is as if each realization of the noise is a distinct world, in the many-worlds sense.

We can look for phenomenological consequences of the fluctuations in the geometry.
Consider Galileo's experiment, in which one drops a massive particle from rest starting at
some height $z_0$ and then measures the time taken for it to reach the ground at $z = 0$. Classically, the Newtonian trajectory is $z_{\rm cl}(t) = z_0 - \frac{1}{2} gt^2$ and the arrival time is $\tau = \sqrt{2z_0/g}$. Let us calculate the effect of quantum gravitational fluctuations on free-fall. To do so, we first neglect all the other corrections to Newton's law, (\ref{eom}), so that our equation reduces to $\ddot{z} =  -g + \ddot{N}_{\Psi} z$. Regarding the quantum fluctuation in the trajectory as a small perturbation over the classical trajectory, we write $z(t) \approx z_{\rm cl}(t) + z_{\rm q} (t)$. Then
\be
\ddot{z}_{\rm q}(t) \approx \ddot{N}_{\Psi} z_{\rm cl}(t).
\ee
Integrating twice, we find
\be
{z}_{\rm q}(t) \approx N_{\Psi} z_{\rm cl}(t) +  2 \int^t N_{\Psi} (t') gt' dt' - \int^t
dt' \int^{t'} N_{\Psi} (t'') g dt''.
\ee
Since $N_{\Psi}$ takes values from a Gaussian distribution with zero mean, (\ref{funcgauss}), we have $\langle \langle N_{\Psi}\rangle \rangle= 0$, where $\langle \langle { \, \cdot \, } \rangle \rangle$ denotes a statistical average over the distribution for $N_\Psi$, and hence
\be
\langle \langle z_{\rm q} (t) \rangle \rangle = 0 \; .
\ee
That is, the average trajectory is the classical trajectory, $\langle \langle z(t) \rangle \rangle = z_{\rm cl} (t)$. We can also calculate the variance. Let us evaluate the variance in the height at the time $\tau = \sqrt{2z_0/g}$, the classical free-fall time at which $z_{\rm cl} (\tau) = 0$:
\begin{eqnarray}
\langle \langle z (\tau) z(\tau) \rangle \rangle & = & 4 g^2 \int^{\tau}  \! t' dt' \int^{\tau} \! \langle \langle N_{\Psi} (t') N_{\Psi} (\tilde{t}) \rangle \rangle \tilde{t} d \tilde{t} - 4 g^2 \int^{\tau} dt' \! \int^{t'}  dt'' \! \int^{\tau} \! \langle \langle N_{\Psi} (t'') N_{\Psi} (\tilde{t}) \rangle \rangle \tilde{t} d \tilde{t} \nonumber \\
& & + g^2 \int^{\tau} \! dt' \int^{t'} \!  dt''
 \int^{\tau} \! d\tilde{t} \int^{\tilde{t}} \! d \tilde{\tilde{t}} \langle \langle N_{\Psi} (t'') N_{\Psi} (\tilde{\tilde{t}}) \rangle \rangle .
\label{variance}
\end{eqnarray}
But terms of the type $\langle \langle N_{\Psi}(t) N_{\Psi}(t') \rangle \rangle$ are just the auto-correlation function $A_{\Psi}(t,t')$ in the gravitational field state $|\Psi \rangle$. For the vacuum state, this is given by (\ref{A0}). The integral over $\omega$ is divergent and we will need to regulate both the upper and the lower limits. Two natural cut-offs are $\omega_{\rm max} = \frac{2 \pi c}{z_0}$ and $\omega_{\rm min} = \frac{2 \pi}{\tau}$. Intuitively, the upper limit comes from our use of the dipole approximation; indeed, the same limit shows up for the sensitivity of gravitational wave detectors. The lower limit arises because our calculation is done over a time $\tau$. Inserting (\ref{A0}) into (\ref{variance}), we find
\be
\langle \langle z(\tau) z(\tau) \rangle \rangle \approx \frac{4g^2 \hbar G}{\pi c^5} \left . \left [\frac{1}{\omega^2} - 3 \frac{\cos (\omega \tau))}{\omega^2} - 3 \tau \frac{\sin (\omega \tau)}{\omega} + 5 \tau^2 \ln \omega - \tau^2 {\rm Ci} (\omega \tau) \right ] \right |_{\omega_{\rm min}}^{\omega_{\rm max}},
\ee
where the special function is given by ${\rm Ci}(x) = - \int_x^{\infty} dt (\cos t)/t$. We find roughly that, upto constants of order unity,
\be
\langle \langle z(\tau) z(\tau) \rangle \rangle \sim \frac{g^2 \hbar G \tau^2}{c^5} \ln \left ( \frac{\omega_{\rm max}}{\omega_{\rm min}} \right ) = \frac{g^2 \hbar G \tau^2}{c^5} \ln \left ( \frac{c \tau}{z_0} \right ) = \frac{g z_0 \hbar G}{c^5} \ln \left ( \frac{2 c^2}{g z_0} \right ).
\ee
Taking the square root yields the standard deviation in the height at the classical arrival time, $\tau$. Finally, we can approximate the spread in arrival times by dividing the standard deviation in the height by the average velocity at time $\tau$, which is $|\dot{z}_{\rm cl} (\tau)| = g \tau$. This scales as $\Delta \tau \propto \sqrt{\ln \frac{2c^2}{gz_0}}$, which varies only by order unity even over many orders in $z_0$. For $z_0$ of a meter we find the standard deviation in the arrival times to be about $10^{-43} s$, or about a Planck time. When the gravitational state is in a squeezed state, however, the auto-correlation function is enhanced by a factor of $\sim e^{2r}$ where $r$ is the squeezing parameter.  Thus the standard deviation in arrival times would also be enhanced exponentially by a factor of $e^r$. It would, however, take a very large $r$ to make the variation in arrival times observable.

\bigskip
\noindent
{\bf Acknowledgments}\\
\noindent
We thank Matthew Baumgart, Yanbei Chen, Rich Lebed, Frank Wilczek, and George Zahariade for helpful conversations. MP is supported in part by Heising-Simons Foundation grant  2021-2818.

\bibliography{newtonv12inspire}
\end{document}